\documentclass[journal=jacsat,manuscript=article]{achemso}

\usepackage[version=3]{mhchem} 
\usepackage[english]{babel}
\usepackage[utf8]{inputenc}
\usepackage{comment}
\usepackage{adjustbox}
\usepackage[pdftex, pdftitle={Article}, pdfauthor={Author}]{hyperref} 
\usepackage{comment}

\author{Engin Torun}
\affiliation{Department of Physics and Materials Science, University of Luxembourg, 162a avenue de la Faïencerie, L-1511 Luxembourg, Luxembourg}
 \altaffiliation{Contributed equally to this work}
 \email{engntr@gmail.com}
\author{Fulvio Paleari}
\affiliation{CNR-NANO, Via Campi 213a, 41125 Modena, Italy}
 \altaffiliation{Contributed equally to this work}
\email{fulvio.paleari@nano.cnr.it}
\author{Milorad V. Milo\v{s}evi\'{c}}%
\affiliation{Department of Physics \& NANOlab Center of Excellence, University of Antwerp, Groenenborgerlaan 171, B-2020 Antwerp, Belgium}
\author{Ludger Wirtz}
\affiliation{Department of Physics and Materials Science, University of Luxembourg, 162a avenue de la Faïencerie, L-1511 Luxembourg, Luxembourg}
\author{Cem Sevik}
\email{csevik@eskisehir.edu.tr}
\affiliation{Department of Mechanical Engineering, Faculty of Engineering, Eskisehir Technical University, 26555 Eskisehir, Turkey}
\alsoaffiliation{Department of Physics \& NANOlab Center of Excellence, University of Antwerp, Groenenborgerlaan 171, B-2020 Antwerp, Belgium}

 \title
  {Intrinsic control of interlayer exciton generation rate in van der Waals materials via Janus layers}

\abbreviations{IR,NMR,UV}

\begin{document}

\begin{abstract}
We demonstrate the possibility of engineering the optical properties of transition metal dichalcogenide heterobilayers when one of the constitutive layers has a Janus structure. This has important consequences for the charge separation efficiency. We investigate different MoS$_2$@Janus layer combinations using first-principles methods including electron-hole interactions (excitons) and exciton-phonon coupling. The direction of the intrinsic electric field from the Janus layer modifies the electronic band alignments and, consequently, the energy separation between interlayer exciton states -- which usually have a very low oscillator strength and hence are almost dark in absorption -- and bright in-plane excitons. We find that in-plane lattice vibrations strongly couple the two states, so that exciton-phonon scattering may be a viable generation mechanism for interlayer excitons upon light absorption. In particular, in the case of MoS$_2$@WSSe, the energy separation of the low-lying interlayer exciton from the in-plane exciton is resonant with the transverse optical phonon modes (40 meV). We thus identify this heterobilayer as a prime candidate for efficient electron-hole pair generation with efficient charge carrier separation.
\end{abstract}

\section{Introduction}
The weak dielectric screening of two-dimensional semiconductors allows for the formation of strongly bound electron-hole pairs (excitons) upon light absorption, leading to peculiar optical properties such as discrete excitonic peaks with strong photoluminescence response and layer-dependent exciton modulation~\cite{PhysRevLett.113.026803,PhysRevLett.105.136805,doi:10.1021/nl903868w,review1}. In this regard, two-dimensional transition metal dichalcogenides (TMDs) are exemplary since, due to the quasi 2D confinement and weak dielectric screening, they host excitons with binding energies of hundreds of meV~\cite{Molina-Sanchez2013Jul,PhysRevLett.111.216805,PhysRevLett.113.076802,RevModPhys.90.021001}. Therefore, this class of materials represents an important testbed for the physics of light-matter interaction~\cite{RevModPhys.90.021001}, as evidenced by a wealth of observed phenomena including exciton/polariton condensation~\cite{20.500.11850/122168,2017Sci3581314K,2016PhRvB93e4510C}, the realization of advanced optoelectronic and nanophotonic devices~\cite{adfm201900040,adma201802687,2018NatRM338A,RossJasonS2014Etel,PMID24608231}, as well as valleytronics.~\cite{2013NatNa8634J,2012PhRvL108s6802X}

Peculiarly, heterobilayer (HBL) structures with different TMD layers generally host interlayer (IL) excitons (where electron and hole forming the exciton reside in different layers) with a static electric dipole moment. Earlier reports showed that the type-II band alignment of the constituent monolayers causes the IL exciton to be the lowest-energy excitation in the HBL absorption spectrum despite having smaller binding energy than in-plane (IP) excitons ~\cite{102a2111K,165306,Deilmann2018,245427,115104}. These IL excitons have a lifetime almost $100$ times longer than the more commonly observed in-plane (intralayer -- IP) excitons --- where electron and hole reside in the same layer~\cite{nl503799t,205423} --- and are also at the forefront of current research. 
For example, their ultrafast formation dynamics is investigated~\cite{ChenHailong2016Ufoi,6b00801}, along with their role in valleytronics~\cite{116055S,7b00175} and charge transfer. 

Ovesen \textit{et al.}~\cite{interlayer_ex} have suggested a scenario for the formation of the IL excitons in HBL TMDs: the excitation of the IP exciton due to light absorption is followed by the tunneling of holes into a finite-momentum state of the opposite layer which can then relax to the ground state of the IL exciton via phonon-scattering. Therefore, the energy-momentum dependence and exciton energy offsets in bilayer structures are crucial for the formation of IL excitons in HBLs. These excited-state features also strongly depend on structural degrees of freedom such as layer separation and stacking. By these means the engineering of optical transition strengths, energies, and selection rules have been previously demonstrated~\cite{b4d69baa99a,articleAta,6198F,7372H,7b00640}, notably by the application of external electric fields\cite{Gao2017Dec,PRA034017,Deilmann2018,Peimyoo2021}.

However, the electric field does not need to be external: the addition of a so-called ``Janus'' layer\cite{acsnano7b03313} can provide a strong intrinsic electric field. 
Janus monolayers have been demonstrated by experimental studies.
Qin \textit{et al.}~\cite{TongayNew} have recently reported an \textit{in situ} growth process which results in Janus TMD monolayers with high structural and optical quality.
A first theoretical prediction of the impact of the intrinsic electric field, with a possible reordering of IP and IL excitons, was done recently for Janus-bilayers\cite{Zhang2021}. 
In this paper, we propose that combining a non-Janus TMD monolayer with a Janus monolayer -- thus creating a TMD@JTMD heterostructure -- allows for reliable tuning of the relative order of the IP and IL excitons, in turn leading to an efficient pathway for IP-to-IL conversion via exciton-phonon scattering.
Experimentally, Trivedi \textit{et al.}~\cite{adma202006320} have already shown the controllable room temperature fabrication and optical characterization of TMD@JTMD and JTMD@JTMD HBL crystals. The tuning of the intrinsic net out-of-plane electric dipole moment in these materials, with the large corresponding piezoelectric effect~\cite{acsnano7b03313}, is also promising for light-energy/electricity interconversions and valley-contrasting physics~\cite{acsjpcc9b09097,jpclett8b01625,C8TA08407F,acsjpcc9b02657}.

\begin{figure}[htb!]
\centering
\includegraphics[width=0.4\textwidth]{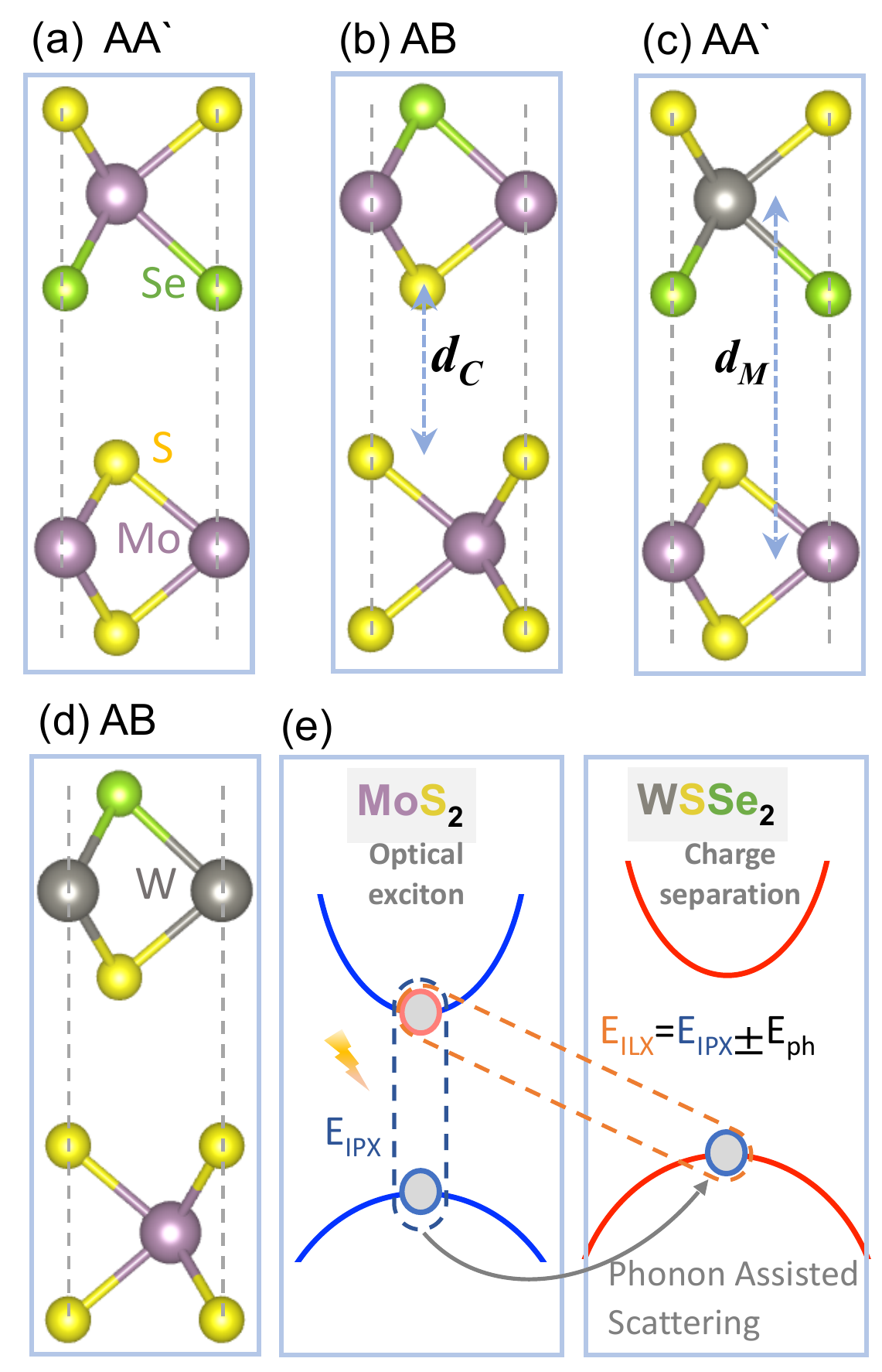}
\caption{(a) to (d): Schematic representation of the investigated TMD@JTMD heterobilayer structures (only the most stable configurations are reported). The TMD layer (bottom) is always MoS$_2$. The JTMD layer (top) is either MoSSe or WSSe, while the TMD@JTMD interface can be either S-S or S-Se. Moreover, the two layers can be stacked in different configurations denoted AA$^\prime$ and AB. (e): Schematic representation of the phonon-assisted IP $\rightarrow$ IL exciton transition in the case of MoS$_2$@WSSe.}
\label{fig1}
\end{figure}

More specifically, we investigate the optoelectronic properties of MoS$_2$@MoSSe, MoS$_2$@MoSeS, MoS$_2$@WSSe, and MoS$_2$@WSeS (as shown in Fig.~\ref{fig1}a-d) using first-principles, many-body perturbation theory techniques including quasiparticle corrections, electron-hole interactions and exciton-phonon coupling~\cite{50005082,RevModPhys73515,MARINI20091392, Sangalli_2019,HEDIN19701,RevModPhys74601,PhysRevB.73.233103,Cudazzo2020}. We demonstrate that the direction of the intrinsic electric field polarization of the Janus layer changes both the band alignments and the energy separation of the lowest-energy IL and IP excitonic levels. In addition, we calculate the scattering strength of the IP-to-IL exciton transitions mediated by optical phonons at the zone center, in order to address the generation mechanisms of the charge-separated and long-lived IL excitons as schematically explained in Fig.~\ref{fig1}e.

\section{Results and discussion}
\subsection{Structural properties and stability with respect to stacking order and Janus layer polarization}

\begin{table}
\footnotesize
\caption{The calculated lattice constants ($a_0$), interlayer separation of metal ($d_M$) and chalcogen ($d_C$) atoms, total energy with respect to minimum energy stacking (E$_{tot}$), electronic direct/indirect band gap at PBE (E$^{DFT}_{gap}$) and  $G_0 W_0$ level (E$^{GW}_{gap}$), inter-layer (IL) and in-plane (IP) exciton energies of the HBLs.}
\begin{adjustbox}{width={\textwidth},totalheight={\textheight},keepaspectratio}
\begin{tabular}{ccccccccc}
HLB&ST&$a_0$&$d_M$/$d_C$&E$_{tot}$&E$^{DFT}_{gap}$ dir./ind.&E$^{GW}_{gap}$ dir./ind. &$\mathbf{IL}$ & $\mathbf{IP}$ \\
&&(\AA)&(\AA)&(meV/atom)&(eV)&(eV)&(eV)&(eV)\\\hline
             &AB&3.22&6.33/3.14& 0.14&0.87/0.82&1.85/1.82&&\\
MoS$_2$@MoSeS&AA&3.22&7.01/3.75&11.48&0.80/0.90&1.79/1.89&&\\
            &\textbf{AA${'}$}&\textbf{3.22}&\textbf{6.41/3.15}&\textbf{0.00}&\textbf{0.83/0.81}&\textbf{1.80/1.81}&\textbf{1.38}& \textbf{1.82}\\\hline
            &\textbf{AB}&\textbf{3.23}&\textbf{6.15/3.06}& \textbf{0.00}&\textbf{1.44/0.93}&\textbf{2.44/1.84}&\textbf{1.90}&\textbf{1.83}\\
MoS$_2$@MoSSe&AA&3.22&6.79/3.70& 9.65&1.35/1.14&2.38/2.08&&\\
            &AA${'}$&3.22&6.18/3.10& 0.19&1.39/0.94&2.39/1.84&&\\\hline       
            &AB&3.22&6.35/3.09& 0.19&0.55/0.67&1.62/1.71&&\\
MoS$_2$@WSeS&AA&3.22&6.97/3.70&13.79&0.46/0.73&1.55/1.75&&\\
            &\textbf{AA${'}$}&\textbf{3.22}&\textbf{6.37/3.10}&\textbf{0.00}&\textbf{0.51/0.65}&\textbf{1.57/1.69} & \textbf{1.15} & \textbf{1.83}\\\hline
            &\textbf{AB}&\textbf{3.22}&\textbf{6.08/3.00}& \textbf{0.00}&\textbf{1.12/0.82}&\textbf{2.20/1.81} & \textbf{1.78} & \textbf{1.83}\\
MoS$_2$@WSSe&AA&3.22&6.73/3.64&12.61&1.00/1.04&2.11/2.03&&\\
       &AA${'}$&3.22&6.12/3.03& 0.29&1.06/0.83&2.14/1.82&&\\
\end{tabular}
\label{table1}
\end{adjustbox}
\end{table}
The calculated results for the geometric and electronic properties are reported in Table~\ref{table1}. The obtained in-plane lattice parameters are in excellent agreement with the reported values\cite{C8CP02802H,C7TC05225A} and do not depend on the stacking order. However, the out-of-plane structural parameters $d_{C}$ (interlayer distance between closest chalcogen layers) and $d_{M}$ (interlayer distance between metal atom layers) change with the stacking type. These stacking-dependent structural changes do not remarkably affect electronic structures, as seen in the calculated band gaps and exciton energies also reported in Table~\ref{table1}. Among all the considered systems (see Supplementary Information~\cite{supp} for more details), the difference in total energies between the various structures is of the order of a few meV, with the so-called AA-stacked structures clearly being less energetically favored due to the in-line chalcogen-chalcogen interlayer interaction. Therefore, within the limit of computational error, either the AB or AA$^{\prime}$ stackings can lead to stable structures depending on the orientation of the Janus layer.

At variance with the stacking order, the polarization direction of the Janus layer (namely S-Mo-S$\leftrightarrow$S-Metal-Se or S-Mo-S$\leftrightarrow$Se-Metal-S, with a different interaction at the layers interface) has a notable effect on both structural and electronic properties. In particular, the energy difference between direct and indirect band gaps, as well as the exciton binding energies, change remarkably. For instance, MoS$_2$@MoSSe(WSSe) HBLs are distinctly indirect band gap materials, whereas MoS$_2$@MoSeS(WSSe) are direct band gap materials within the limits of thermal fluctuations. 
\section{Excitonic properties}
\begin{figure}[htb!]
\centering
\includegraphics[width=1.0\textwidth]{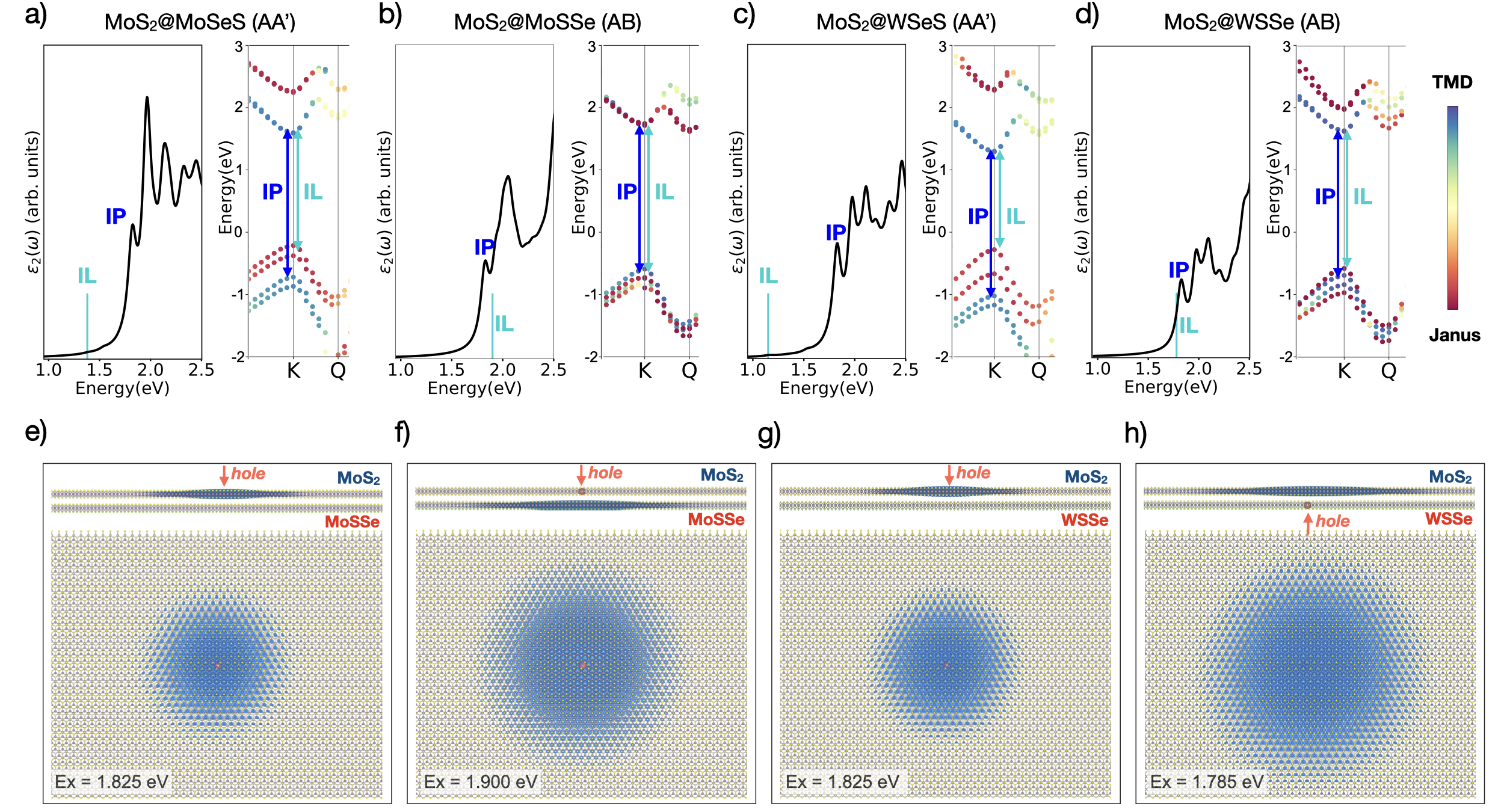}
\caption{Excitonic properties of TMD@JTMD HBLs. Top panels (a to d): optical absorption spectra and quasiparticle band structures (zoomed-in to the vicinity of the \textit{K} point in the BZ, see the supplementary materials for the full band structures) of the investigated systems. The absorption spectrum is proportional to the imaginary part of the excitonic dielectric function $\varepsilon_2$, which is the plotted quantity on the left panels. The band color (right panels) represents the projections of the electronic wave function onto its atomic orbitals components localized either in the JMTD or in the TMD layers (blue: mostly TMD state, red: mostly Janus state). The calculations are performed on the stacking configurations with minimal energy and are: (a) MoS$_2$@MoSeS in AA$^\prime$ stacking, (b) MoS$_2$@MoSSe in AB stacking, (c) MoS$_2$@WSeS in AA$^\prime$ stacking and (d) MoS$_2$@WSSe in AB stacking. The most important single-particle transitions forming the IL (Janus $\leftrightarrow$ MoS$_2$), IP (MoS$_2$ $\leftrightarrow$ MoS$_2$) excitons are labeled in the band plots, while the energy of the resulting exciton states is emphasized in the absorption plots. 
Bottom panels (e to h): Wave-function intensity plot in real space for excitons corresponding to (e) IP in MoS$_2$@MoSSe, (f) IL in MoS$_2$@MoSSe, (g) IP in MoS$_2$@WSSe, and (h) IL in MoS$_2$@WSSe. The plot is obtained by fixing the hole in a position consistent with its band orbital character and plotting the resulting electron distribution.}
\label{fig2}
\end{figure}

The results for quasiparticle band structures (in the G$_0$W$_0$ approximation) and the absorption spectra including electron-hole interactions for AA$^{'}$ and AB stacked HBLs are displayed in Fig. ~\ref{fig2}(a-d) (left: excitonic optical absorption spectrum; right: quasiparticle band structure in the vicinity of the \textit{K} point of the hexagonal Brillouin zone (BZ); see Supplementary Information~\cite{supp} for all the considered materials). The band structures include the projections of the electronic wave functions onto atomic orbitals localized on the constituent layers: the red and blue colors represent electronic states mostly localized on the Janus and on the TMD monolayer, respectively. Electron-hole interactions were included via the solution of the Bethe-Salpeter equation (BSE) from first principles.\cite{Martin2016}

The investigated structures differ in stacking geometry (AA$^{\prime}$ vs AB), transition metal in the Janus layer (Mo vs W) and orientation of the Janus layer, controlling the polarization direction (S$\leftrightarrow$S vs S$\leftrightarrow$Se interface between the two layers).
All the HBLs present some common optoelectronic features, as well as some notable differences.
The electronic states around the $K$ points in the quasiparticle band structures are completely confined on one of the two layers. In particular, the four highest valence states comprise two pairs of spin-orbit-split bands, with one pair localised on the TMD layer (blue color) and the other pair on the Janus layer (red color). The same is true for the four lowest conduction states.
In addition, the minimum energy transition at the $K$ point always connects a valence state on the Janus layer to a conduction state on the TMD one.
As a consequence of the band structure shape, the low-energy region of the excitonic absorption spectra is always dominated by the IP and IL excitons originated from single-particle transitions in the bands around $K$.
Another common excitonic feature is that the low-lying IP states always originate from electronic transition between the TMD valence and conduction bands (blue to blue in the band graph), therefore they are always localised on the TMD layer, as shown in two examples in Fig. \ref{fig2}e-g.


The four HBLs however differ remarkably with respect to the nature of the quasiparticle band gap, which is direct at $K$ for the AA$^\prime$ stackings with S-Se interface (Fig. ~\ref{fig2}(a) and (c)), while indirect from $\Gamma$ to $K$ for the AB stackings with S-S interface (Fig. ~\ref{fig2}(b) and (d)).
Another important difference is due to the effect of the intrinsic electric field arising from the Janus monolayer, which shifts the TMD bands with respect to the Janus ones in both valence and conduction states. The direction of the shift depends on the direction of the field, i.e., on the orientation of the Janus layer.
More specifically, in the HBLs with S/Se interface the TMD bands are shifted \textit{away} from the Janus ones, leading to a substantial energy separation. When the Janus dipole orientation is reversed, instead, the TMD bands are shifted \textit{closer} to the Janus ones becoming almost degenerate with them.
These changes at the quasiparticle level determine the most important optical feature for charge transfer efficiency: the energy difference -- and thus likelihood of transition -- between the IL and IP excitonic states, which varies considerably between the systems.
Intriguingly, the localization of the IL excitons changes depending on the stacking and JTMD layer types. In MoS$_2$@MoSSe, the exciton is formed by electronic transitions from the TMD layer to the JTMD one (blue to red in the band graph), while in all other cases the hole is on the JTMD layer and the electron on the TMD one (red to blue in the band graph). This difference in spatial localization can be seen in Fig.~\ref{fig2}f-g.

\textit{MoS$_2$@MoSeS (AA$^{\prime}$ stacking, S-Se interface).}
The geometry of this system is represented in Fig.~\ref{fig1}(a), the quasiparticle/BSE 
 results in Fig.~\ref{fig2}(a).
This HBL is a direct gap material with a gap of 1.80 eV. The indirect band gap sits 0.01 eV above. Note that at the DFT-PBE level 
 this energy ordering is reversed as seen in Table~\ref{table1}, which shows that the quasiparticle correction to the Kohn-Sham energies is not just a rigid shift of the bands owing to the $k$-dependence of the $GW$ self-energy.  
The ordering of the bands at the $K$ point indicates a type-II character for this HBL at the quasiparticle level, as previously observed for MoS$_2$@WS$_2$ and MoSe$_2$@WSe$_2$ HBLs.\cite{245427,102a2111K,165306,115104} The energy of the IL exciton which forms via the transitions from the valence band maximum (VBM) to conduction band minimum (CBM) is 1.38 eV. The energy difference between the lowest energy IL and the first in-plane exciton, IP, is 0.44 eV, which is too large for a one-phonon scattering process to enable the direct transitions from the optically excited IP to the IL state. 

\textit{MoS$_2$@MoSSe (AB stacking, S-S interface).}
The geometry of this system is represented in Fig.~\ref{fig1}(b) and the quasiparticle/BSE results are shown in Fig.~\ref{fig2}(b).
Contrary to the previous case, here we have an indirect gap semiconductor at both the DFT-PBE (0.93 eV) and G$_0$W$_0$ (1.84 eV), respectively. Because of the direction of the intrinsic electric field, the TMD and Janus bands overlap at the CBM and have very similar band energies at the VBM. This leads to  a situation where the electron-hole interaction strength of the IL and IP excitons become the determining factor for the energy ordering of these excitons in the absorption spectrum. In MoS$_2$@WS$_2$ and MoSe$_2$@WSe$_2$ HBLs it has been shown that the binding energies\footnote{Here, the binding energy of an IP (IL) exciton is defined as the difference between the exciton energy and the lowest-energy single-particle IP (IL) transition.} of IL excitons are approximately 100 meV lower than the ones of IP excitons ~\cite{245427}. The case of the MoS$_2$@MoSSe HBL, here, is similar: the IP exciton (1.83 eV) has approximately 70 meV higher binding energy than the first IL exciton (1.90 eV) as reported in Table~\ref{table1}. Due to this, the IP exciton automatically becomes the lowest energy exciton in the absorption spectrum as shown in Fig.~\ref{fig1}(b). 
The spatial distribution of the excitonic wave functions for this system is shown in Fig. \ref{fig2}e and f for the IP and IL states, respectively.
The energy separation between the IP and the IL exciton is around $70$ meV. This is not ideal because the charge-separated state is not energetically favored, and even a thermal population of the IL exciton would be very tiny.

\textit{MoS$_2$@WSeS (AA$^{\prime}$ stacking, S-Se interface).}
The geometry of this system is represented in Fig.~\ref{fig1}(c) and the quasiparticle/BSE results are shown in Fig.~\ref{fig2}(c).
Similar to the Mo case, this is a direct band gap material with a quasiparticle gap of 1.57 eV (see Table\ref{table1}). Again, orbital projections indicate the type-II character of the electronic bands in this HBL. The SOC splitting of the Janus bands, shown in red, is more pronounced than in the Mo case due to the presence of the heavier W atoms. 
Here the energy difference between the IL exciton (1.15 eV) and the lowest-energy IP exciton (1.83 eV) is 0.65 eV, even larger than in the Mo case with the same stacking and Janus orientation, due to the more substantial SOC. This again rules out efficient IL exciton generation from the IP states via first-order phonon-assisted conversion processes, with incoherent exciton scattering (i.e., relaxation dynamics\cite{interlayer_ex,Chen2022}) likely being the most important mechanism.

\textit{MoS$_2$@WSSe (AB stacking, S-S interface).}
The geometry of this system is represented in Fig.~\ref{fig1}(d) and the quasiparticle/BSE results are shown in Fig.~\ref{fig2}(d).
This HBL has an indirect quasiparticle band gap of 1.81 eV.
At the VBM, the TMD bands (blue) are squeezed in between SOC-split Janus bands (red). Compared to the previous W-based HBL, the opposite intrinsic dipole moment from the Janus layer shifts the bands of the MoS$_2$ layer so that the bands localized on the two layers are energetically very close to each other just as in the case of the Mo-based system with the same geometry. The lowest-lying IL state (1.78 eV) is just 40 meV below the IP exciton as shown in Fig.~\ref{fig2}(d), while the real-space excitonic wavefunctions are represented in Fig. \ref{fig2}g-h.  
We will see in the following section that this small energy difference will drastically improve the efficiency of the exciton-phonon scattering channel between the two excitonic states and, hence, the charge carrier separation efficiency of this HBL.  

\section{Exciton-phonon coupling strengths}
We now focus on MoS$_2$@WSSe (AB stacking), the HBL where the lowest-bound intralayer and interlayer excitons have a very small energy separation ($40$ meV), lower than the Debye energy ($56$ meV). 
This means that phonon-mediated charge separation, i.e., excitonic intralayer-interlayer (IP-IL) scattering might be very efficient, due to two concurring mechanisms.
First, the exciton relaxation dynamics suggests that after the higher-energy intralayer exciton is photoexcited, incoherent scatterings mediated by low-momentum acoustic phonons will quickly transfer the carriers to the lower-energy interlayer state (the description of out-of-equilibrium carrier dynamics is beyond the scope of this paper).
Second, direct intralayer-interlayer scatterings mediated by optical phonons at vanishing momenta will also be permitted and may play an important role.
We quantitatively analyze the latter mechanism by computing the exciton-phonon coupling matrix elements $\mathcal{G}$ at zero exciton and phonon momenta for this system\cite{Reichardt2020Aug,Chen2020,Cudazzo2020,Antonius2022}:   
\begin{equation}\label{eq:exc-ph}
  \mathcal{G}_{\alpha\beta}^{\mu}=\sum_{vck}\left[\sum_{v^\prime} \left(A^{cv^\prime k}_\beta\right)^* g_{vv^\prime k}^\mu A^{cvk}_\alpha - \sum_{c^\prime} \left(A^{c^\prime v k}_\beta\right)^* g_{c^\prime ck}^\mu A^{cvk}_\alpha\right]
\end{equation}

\begin{figure}
\centering
\includegraphics[width=\textwidth]{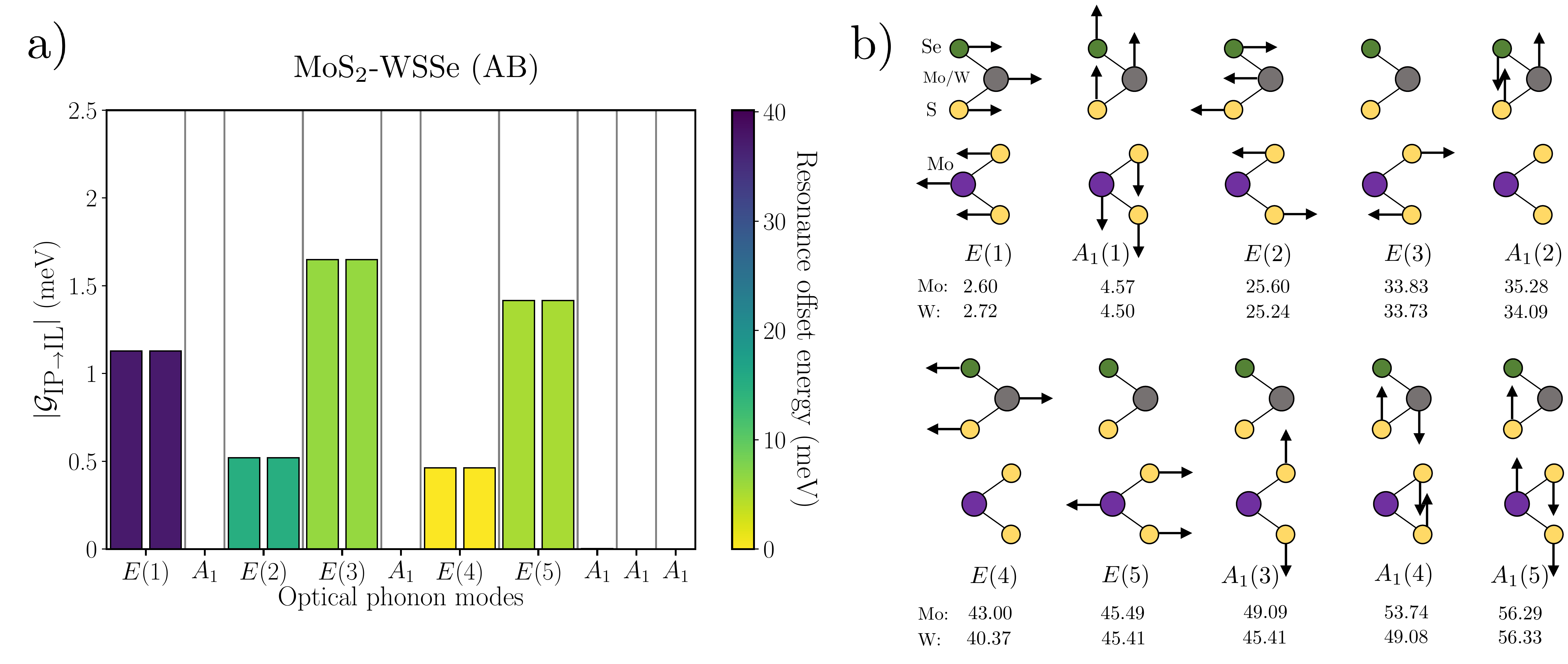}
\caption{(a) Exciton-phonon coupling strengths, $|\mathcal{G}|$ (see Eq. \eqref{eq:exc-ph}) for the intralayer (IP) to interlayer (IL) exciton scattering per phonon mode in MoS$_2$@WSSe (AB stacking, see Fig. \ref{fig2}[d]). The colors report the resonance energy offset values (see text). (b) Oscillation patterns of the zone-center optical phonon modes for MoS$_2$@MoSSe (AB stacking) and MoS$_2$@WSSe (AB stacking). The arrow lengths are not scaled by the ionic masses, but they are set to zero for oscillations one order of magnitude weaker than the rest. The phonon energies associated with each mode are reported, in meV, for both systems.}
\label{fig3}
\end{figure}

Here we assume that excitons can be approximately described as well-defined quasiparticle excitations with bosonic character.\cite{paleari2022}
In Eq. \eqref{eq:exc-ph}, $\alpha$ and $\beta$ are the indices of the exciton states involved in a scattering mediated by a phonon mode $\mu$. The exciton eigenvectors $A^{cvk}$ are expressed in terms of single-particle transitions at wave vector $k$ from a valence band $v$ to a conduction band state $c$. They represent the excitonic wave function and result from the solution of the BSE. The $g_{c'ck}^\mu$ ($g_{vv'k}^\mu$) are the \textit{electron}-phonon coupling matrix elements, obtained from a density functional perturbation theory (DFPT) calculation and representing  the scattering amplitude from a conduction band state $c$ (valence band state $v'$) at wave-vector $k$ into another state $c'$ ($v$) at the same wave vector, via absorption/emission of a phonon mode $\mu$.
Thus, the values of $|\mathcal{G_{\alpha\beta}^\mu}|$ represent the coupling strengths of excitonic transitions $\alpha \rightarrow \beta$ via absorption/emission of phonon mode $\mu$.
The calculated values for the IL-IP scattering are displayed in Fig. \ref{fig3}a, showing that out of the $10$ distinct optical phonon modes present in these material, all those with atoms oscillating in the layer plane may couple the two excitons and represent possible scattering channels. These are doubly degenerate modes with $E$ symmetry. The coupling is instead forbidden for the out-of-plane phonons.
This confirms that an IP exciton localised on one layer may transfer its carriers to the lower-lying IL one with the help of ionic oscillations in that layer. In the case of MoS$_2$@WSSe, the coupling strengths vary between $0.5$ and $2$ meV. 
A scheme of the oscillation patterns\cite{molinasanchez2015} is provided in Fig. \ref{fig3}b.

The other factor affecting scattering probabilities is energy conservation. We define the resonance offset energy (ROE) as $\Delta E_{\alpha,\beta}^\mu =|E_{\alpha}-E_{\mu}-E_{\beta}|$ ($E_{\mu}$ being the phonon energy).
The closer $\Delta E_{\alpha,\beta}^\mu$ is to zero, the more likely the transition is to happen. 
Here we consider only the phonon emission case, since it is the dominant contribution with respect to phonon absorption.
The ROE values are shown with a color scale in Fig. \ref{fig3}a.
For MoS$_2$@WSSe, the energy of the fourth $E$ phonon mode is exactly resonant with the excitonic transition at $40.4$ meV. This phonon mode corresponds to ionic oscillation of the Janus layer only. In addition, strongly coupled $E$ modes three and five (mostly the TMD layer moving) are both just $5$ meV from the resonance.
Therefore, our calculations predict that the IP-IL excitonic scattering mediated by optical phonons will be a particularly efficient process for the W-based HBL.

\section{Conclusions}
We have shown that the intrinsic net out-of-plane electric dipole moment of a Janus-type TMD layer has a strong influence on the optoelectronic properties of heterobilayer systems in which it is used. 
In particular, our first-principles analysis on MoS$_2$@MoSSe, MoS$_2$@MoSeS, MoS$_2$@WSSe, and MoS$_2$@WSeS demonstrates that the polarization direction of the Janus layer can be used to tune the dynamics of excitons -- most notably by altering the energy separation between interlayer and in-plane excitonic states -- without the use of external fields.
Surprisingly, for MoS$_2$@WSSe the calculated energy difference is exactly resonant with in-plane optical phonon modes. Moreover, the calculated zero-momentum exciton-phonon couplings point to efficient in-plane to interlayer excitonic scattering mediated by optical phonons, again with MoS$_2$@WSSe being the prime candidate for this mechanism.

It is important to note, however, that this system also has an indirect band gap: hence, new low-lying dark excitonic states at finite momentum may be important, introducing an additional possible pathway for exciton dynamics which could be detrimental to the intra- to interlayer conversion rate. This opens a relevant future avenue of investigation theoretically, experimentally, and from a materials design perspective.
Theoretically, the next step is computing the exciton-phonon couplings at finite momenta and simulating the excitonic relaxation dynamics in order to compare the transition rates of the various competing mechanisms. Experimentally, these quantities can be measured by photoluminescence and time-resolved, pump-and-probe studies. In materials design, it would be important to target excited-state properties both microscopically and at the excitonic level, rather than macroscopic optical properties at the single-particle level, in order to obtain candidate systems more effectively.

In conclusion, our results clearly support the use of Janus materials in layer engineering to boost the generation rate of long-lived interlayer excitons in heterobilayer TMD crystals. These excitons are obtained by phonon-assisted conversion of optically excited intralayer states.

\section{Methods}
The single-particle wave functions and corresponding energies (DFT step) are obtained from density functional theory as implemented in the Quantum ESPRESSO code (QE)~\cite{50005082} using Perdew-Burke-Ernzerhof (PBE)~\cite{773865} norm-conserving, fully relativistic pseudopotentials in the generalized gradient approximation (GGA)~\cite{085117}.
These were generated by the PseudoDojo project~\cite{VANSETTEN201839}. 
The plane-wave energy cutoff, vacuum separation between periodic repetitions of the simulation supercell, and $k$-grid sampling are,  120~Ry, 55~a.u. and 42$\times$42$\times$1 ($\Gamma$ centered), respectively. We adopted Grimme’s dispersion correction (labeled as Grimme-D2 in QE)\cite{Grimme_1,Grimme_2} in order to take van der Waals (vdW) interactions into account.
The Many-Body Perturbation Theory calculations\cite{Martin2016}, performed on top of the DFT results, were conducted with the YAMBO code.\cite{MARINI20091392, Sangalli_2019}
The G$_0$W$_0$\cite{HEDIN19701,RevModPhys74601} corrections to the single-particle eigenvalues were computed with the plasmon-pole approximation for the dynamical electronic screening.
The direct and indirect band gaps were converged with the 42 $\times$ 42 $\times$ 1 $k$-grid mesh (yielding 169 $k$-points in the irreducible Brillouin zone), summing over $600$ and $900$ states for the screening and the Green's functions, respectively.  The corrections were computed for the top $4$ valence bands and the bottom $4$ conduction bands.
The BSE\cite{Martin2016} for excitons was then solved in the Tamm-Dancoff approximation with RPA static screening, which was summed over $300$ bands.
The direct exciton energies and their wave functions were obtained for the first $12000$ excitonic states by using the iterative scheme enabled by the SLEPC library.~\cite{slepc} The Coulomb cutoff (CC) technique was used along the out-of-plane direction to eliminate the long-ranged interactions with the repeated periodic images of the systems in both G$_0$W$_0$ and BSE steps.~\cite{PhysRevB.73.233103} The same computational settings used at the DFT level, albeit with stricter convergence thresholds for the electronic wave functions, were adopted for the $\Gamma$-point calculation of phonon frequencies, eigenvector displacements and electron-phonon coupling matrix elements via Density Functional Perturbation Theory as implemented in the Quantum Espresso Code.~\cite{RevModPhys73515}
A dedicated Coulomb cutoff technique\cite{sohier2017} was employed also in the phonon case.

\begin{suppinfo}
Supplementary Information file includes the schematic representation of considered Janus HBLs, calculated band structures at both PBE and G$_{0}$W$_{0}$ levels, and calculated $\varepsilon_{2}$ with quasi-particle (via Bethe Salpeter Equation) and independent particle approximations.   
\end{suppinfo}

\begin{acknowledgement}
C. S. acknowledges funding by the the Air Force Office of Scientific Research (AFOSR, USA) under award number FA9550-19-1-7048. M. M acknowledges the Research Foundation-Flanders (FWO-Vlaanderen). F.P. acknowledges the European Union project: MaX {\em Materials design at the eXascale} H2020-INFRAEDI-2018-1, grant agreement n. 824143. L.W. acknowledges funding by the Fond National de Recherche, Luxembourg via project INTER/19/ANR/13376969/ACCEPT.
\end{acknowledgement}

\providecommand{\noopsort}[1]{}\providecommand{\singleletter}[1]{#1}%
\providecommand{\latin}[1]{#1}
\makeatletter
\providecommand{\doi}
  {\begingroup\let\do\@makeother\dospecials
  \catcode`\{=1 \catcode`\}=2 \doi@aux}
\providecommand{\doi@aux}[1]{\endgroup\texttt{#1}}
\makeatother
\providecommand*\mcitethebibliography{\thebibliography}
\csname @ifundefined\endcsname{endmcitethebibliography}
  {\let\endmcitethebibliography\endthebibliography}{}

\end{document}